# Self-organization and fractality created by gluconeogenesis in the metabolic process

Valeriy Grytsay

Bogolyubov Institute for Theoretical Physics, 14b, Metrolohichna Str., Kyiv 03680
 (E-mail: vgrytsay@bitp.kiev.ua)

**Abstract.** Within a mathematical model, the process of interaction of the metabolic processes such as glycolysis and gluconeogenesis is studied. As a result of the running of two opposite processes in a cell, the conditions for their interaction and the self-organization in a single dissipative system are created. The reasons for the appearance of autocatalysis in the given system and autocatalytic oscillations are studied. With the help of a phase-parametric diagram, the scenario of their appearance is analyzed. The bifurcations of the doubling of a period and the transition to chaotic oscillations according to the Feigenbaum scenario and the intermittence are determined. The obtained strange attractors are created as a result of the formation of a mixing funnel. Their complete spectra of Lyapunov indices, KS-entropies, "horizons of predictability," and the Lyapunov dimensions of strange attractors are calculated. The conclusions about the reasons for variations of the cyclicity in the given metabolic process, its stability, and the physiological state of a cell are made.
**Keywords:** *Gluconeogenesis, glycolysis, metabolic process, self-organization, fractality, strange attractor, Feigenbaum scenario.*

## 1 Introduction

Gluconeogenesis is a biochemical process of formation of glucose from hydrocarbonless predecessors such as pyruvates, aminoacids, and glycerin. The biosynthesis of glucose runs analogously to glycolysis, but in the reverse direction. Gluconeogenesis is realized by means of the inversion of seven invertible stages of glycolysis. Three remaining stages of glycolysis are exergenous and, therefore, irreversible. They are replaced by three "by-pass reactions" that are thermodynamically gained for the synthesis of glucose. Since gluconeogenesis uses the same invertible reactions, as glycolysis does, its biochemical evolution occurred, apparently, jointly with glycolysis. Maybe, the symbiosis of these biochemical processes arose else in protobionts 3.5 bln years in Earth's oxygenless atmosphere. It can be considered as one of the primary open nonlinear biochemical systems, being far from the equilibrium. The self-organization of the given biochemical system resulted in the appearance of a stable dissipative system independent of other biochemical processes of a primary broth. The directedness of the running of a reaction in it was determined by the energy-gained balance. The organic molecule $ATP$, which was formed as a result of glycolysis, became the principal carrier of the energy consumed in all other biochemical processes. This created the conditions of self-organization of other biochemical processes that used $ATP$ as the input product of a reaction. But if the need in glucose arose in other biochemical processes, the directedness of biochemical reactions in the system was changed by the opposite one. In the course of the subsequent biochemical evolution, the given dissipative system is conserved and is present in cells of all types, which indicates their common prehistory.

Thus, the studies of the reactions of gluconeogenesis are determined in many aspects by the results of studies of glycolysis. The direct sequence of reactions with the known input and output products is studied easier than the reverse one.

The experimental studies of glycolysis discovered autooscillations [1]. In order to explain their origin, a number of mathematical models were developed [2-4]. Sel'kov explained the appearance of those oscillations by the activation of phosphofructokinase by its product. In the Goldbeter--Lefever model, the origin of autooscillations was explained by the allosteric nature of the enzyme. Some other models are available in [5-7].

The present work is based on the mathematical model of glycolysis and gluconeogenesis, which was developed by Professor V.P. Gachok and his coauthors [8-10]. The peculiarity of his model consists in the consideration of the influence of the adeninenucleotide cycle and gluconeogenesis on the phosphofructokinase complex of the given allosteric enzyme. This allowed one to study the effect of these factors as the reason for the appearance of oscillations in glycolysis.

At the present time, this model is improved and studied with the purpose to investigate gluconeogenesis. Some equations were added, and some equations were modified in order to describe the complete closed chain of the metabolic process of glycolysis-gluconeogenesis under anaerobic conditions. The developed complete model allows us to consider glycolysis-gluconeogenesis as a united integral dissipative structure with a positive feedback formed by the transfer of charges with the help of $NAD$. Glycolysis with gluconeogenesis is considered as an open section of the biosystem, which is self-organized by itself at the expense of input and output products of the reaction in a cell, which is a condition of its survival and the evolution. The appearance of an autocatalytic process in the given dissipative structure can be a cause of oscillatory modes in the metabolic process of the whole cell.

Gluconeogenesis occurs in animals, plants, fungi, and microorganisms. Its reactions are identical in all tissues and biological species. Phototrophs transform the products of the own photosynthesis in glucose with the help of gluconeogenesis. Many microorganisms use this process for the production of glucose from a medium, where they live.

The conditions modeled in the present work are established in muscles after an intense physical load and the formation of a large amount of lactic acid in them. As a result of the running of the reverse reaction of gluconeogenesis, it is transformed again in glucose.

## 2 Mathematical Model

The given mathematical model describes glycolysis-gluconeogenesis under anaerobic conditions, whose output product is lactate. At a sufficient level of glucose, the process runs in the direct way. At the deficit of glucose, it runs in the reverse one: lactate is transformed in glucose.

The general scheme of the process of glycolysis-gluconeogenesis is presented in Fig.1. According to it, the mathematical model (1) - (16) is constructed with regard for the mass balance and the enzymatic kinetics.

The equations describe variations in the concentrations of the corresponding metabolites: (1) – lactate $L$; (2) – pyruvate $P$; (3) - 2-phosphoglycerate $\psi_3$; (4) – 3-phosphoglycerate $\psi_2$; (5) - 1,3-diphosphoglycerate ($\psi_1$); (6) - fructose-1,6-diphosphate ($F_2$); (7) – fructose-6-phosphate ($F_1$); (8) – glucose $G$; (9), (10), and (11) - $ATP$, $ADP$, and $AMP$, respectively, form the adeninenucleotide cycle at the phosphorylation; (12) - $R_1$ and (13) - $R_2$ (two active forms of the allosteric enzyme phosphofructokinase; (14) - $T_1$ and (15) - $T_2$ (two inactive forms of the allosteric enzyme phosphofructokinase; (16) - $NAD \cdot H$ (where $NAD \cdot H(t) + NAD^+(t) = M$).

Variations of the concentrations of omitted metabolites have no significant influence on the self-organization of the system and are taken into account in the equations generically. Since glycolysis and gluconeogenesis on seven sections of the metabolic chain are mutually reverse processes, only the coefficients are changed, whereas the system of equations describing glycolysis is conserved [8-10]. The model involves the running of gluconeogenesis on the section: glucose - glucose-6-phosphate. Here in the direct way with the help of the enzyme hexokinase, the catabolism of glucose to glucose-6-phosphate occurs. In the reverse direction with the help of the enzyme glucose-6-phosphatase, glucose is synthesized from glucose-6-phosphate. Thus, the positive feedback is formed on this section.

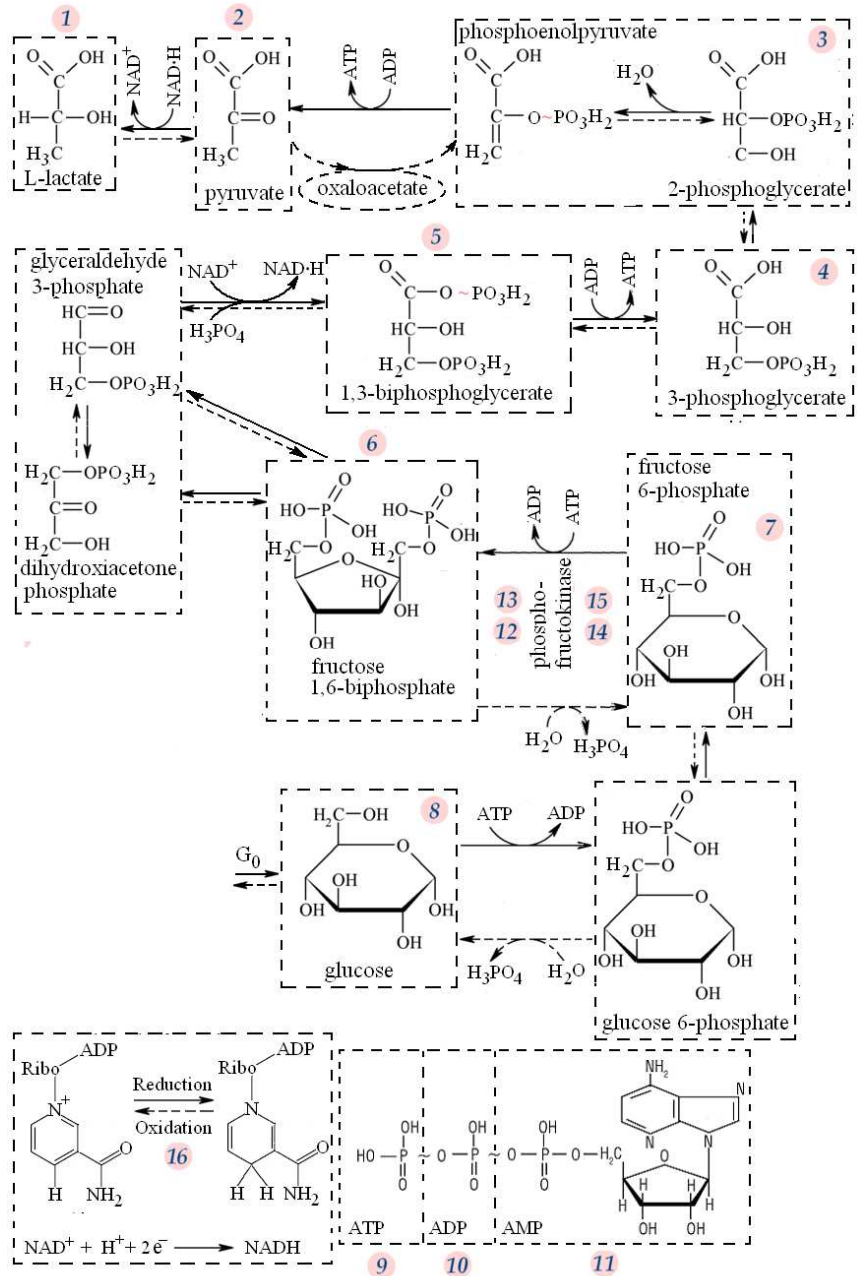

Fig.1. General scheme of the metabolic process of glycolysis-gluconeogenesis.

$$\frac{dL}{dt} = l_7 V(N)V(P) - m_9 \frac{L}{S}, \tag{1}$$

$$\frac{dP}{dt} = l_2 V(\psi_3)V(D) - m_6 \frac{P}{S} - l_7 V(N)V(P), \tag{2}$$

$$\frac{d\psi_3}{dt} = \frac{\psi_2}{S}\frac{m_2}{m_2+\psi_3} - l_2 V(\psi_3)V(D) - m_4\frac{\psi_3}{S}, \quad (3)$$

$$\frac{d\psi_2}{dt} = l_6 V(\psi_1)V(D) - m_8\frac{\psi_2}{S}, \quad (4)$$

$$\frac{d\psi_1}{dt} = \frac{m_5(F_2/S)}{S_1 + m_5(F_2/S)} - l_6 V(\psi_1)V(D) + m_7 V(M-N)V(P), \quad (5)$$

$$\frac{dF_2}{dt} = l_1 V(R_1)V(F_1)V(T) - l_5\frac{1}{1+\gamma A}V(F_2) - m_5\frac{F_2}{S}, \quad (6)$$

$$\frac{dF_1}{dt} = l_8 V(G)V(T) - l_1 V(R_1)V(F_1)V(T) + l_5\frac{1}{1+\gamma A}V(F_2) - m_3\frac{F_1}{S}, \quad (7)$$

$$\frac{dG}{dt} = \frac{G_0}{S}\frac{m_1}{m_1+F_1} - l_8 V(G)V(T), \quad (8)$$

$$\frac{dT}{dt} = l_2 V(\psi_3)V(D) - l_1 V(R_1)V(F_1)V(T) + l_3\frac{A}{\delta+A}V(T) - l_4\frac{T^4}{\beta+T^4} + \\ + l_6 V(\psi_1)V(D) - l_9 V(G)V(T), \quad (9)$$

$$\frac{dD}{dt} = l_1 V(R_1)V(F_1)V(T) - l_2 V(\psi_3)V(D) + 2\cdot l_3\frac{A}{\delta+A}V(T) - \\ - l_6 V(\psi_1)V(D) + l_9 V(G)V(T), \quad (10)$$

$$\frac{dA}{dt} = l_4\frac{T^4}{\beta+T^4} - l_3\frac{A}{\delta+A}V(T), \quad (11)$$

$$\frac{dR_1}{dt} = k_1 T_1 V(F_1^2) + k_3 R_2 V(D^2) - k_5 R_1\frac{T}{1+T+\alpha A} - k_7 R_1 V(T^2), \quad (12)$$

$$\frac{dR_2}{dt} = k_5 R_1\frac{T}{1+T+\alpha A} - k_3 R_2 V(D^2) + k_2 T_2 V(F_1^2) - k_8 R_2 V(T^2), \quad (13)$$

$$\frac{dT_1}{dt} = k_7 R_1 V(T^2) - k_6 T_1\frac{T}{1+T+\alpha A} + k_4 T_2 V(D^2) - k_1 T_1 V(F_1^2), \quad (14)$$

$$\frac{dT_2}{dt} = k_6 T_1\frac{T}{1+T+\alpha A} - k_4 T_2 V(D^2) - k_2 T_2 V(F_1^2) + k_8 R_2 V(T^2), \quad (15)$$

$$\frac{dN}{dt} = -l_7 V(N)V(P) + l_7 V(M-N)V(\psi_1). \quad (16)$$

Here, $V(X) = X/(1+X)$ is the function that describes the adsorption of the enzyme in the region of a local coupling. The variables of the system are dimensionless [8-10]. We take $l_2 = 0.046$; $l_3 = 0.0017$; $l_4 = 0.01334$; $l_5 = 0.3$; $l_6 = 0.001$; $l_7 = 0.01$; $l_8 = 0.0535$; $l_9 = 0.001$; $k_1 = 0.07$; $k_2 = 0.01$; $k_3 = 0.0015$; $k_4 = 0.0005$; $k_5 = 0.05$; $k_6 = 0.005$; $k_7 = 0.03$; $k_8 = 0.005$; $m_1 = 0.3$; $m_2 = 0.15$; $m_3 = 1.6$; $m_4 = 0.0005$; $m_5 = 0.007$; $m_6 = 10$; $m_7 = 0.0001$; $m_8 = 0.0000171$; $m_9 = 0.5$; $G_0 = 18.4$; $L = 0.005$; $S = 1000$; $A = 0.6779$; $M = 0.005$; $S_1 = 150$; $\alpha = 184.5$; $\beta = 250$; $\delta = 0.3$; $\gamma = 79.7$.

In the study of the given mathematical model (1)-(16), we have applied the theories of dissipative structures [11] and nonlinear differential equations [12,13], as well as the methods of mathematical modeling used in author's works [14-34]. In the numerical solution, we applied the Runge--Kutta--Merson method. The accuracy of calculations is $10^{-8}$. The duration for the system to asymptotically approach an attractor is $10^6$.

## 3 The Results of Studies

The mathematical model includes a system of nonlinear differential equations (1)-(16) and describes the open nonlinear biochemical system involving glycolysis and gluconeogenesis. In it, the input and output flows are glucose and lactate. Namely the concentrations of these substances form the direct or reverse way of the dynamics of the metabolic process. Both processes are irreversible and are running in the open nonlinear system, being far from the equilibrium. The presence of the reverse way of gluconeogenesis in the glycolytic system is the reason for the autocatalysis in it. In addition, the whole metabolic process of glycolysis is enveloped by the feedback formed by redox reactions with the transfer of electrons with the help of $NAD$ (16) and the presence of the adeninenucleotide cycle (9) – (11) (Fig.1).

We now study the dependence of the dynamics of the metabolic process of glycolysis-gluconeogenesis on the value of parameter $l_5$ characterizing the activity of gluconeogenesis. The calculations indicate that, as the value of this parameter increases to 0.234, the system passes to the stationary state. As this parameter increases further, the autooscillations of a 1-fold periodic cycle $1 \cdot 2^0$ arise and then, at $l_5 \approx 0.2369$, transit to chaotic ones $- \cdot 2^x$. The analogous behavior of the system is observed at larger values of $l_5$. As the parameter decreases to 0.43, the system stays in a stationary state. If the parameter $l_5$ decreases further, the system gradually transits in a 1-fold periodic cycle $1 \cdot 2^0$, and the region of oscillatory dynamics arises.

Let us consider the oscillatory dynamics of this process. We constructed the phase-parametric diagrams, while $l_5$ varies in the intervals 0.235 – 0.28 and 0.25 – 0.266 (Fig.2,a,b). The diagrams are presented for fructose-6-phosphate $F_1$. We emphasize that the choice of a diagram for the namely given variable is arbitrary. The diagrams of other components are analogous by bifurcations. We want to show that the oscillations on the section fructose-6-phosphate – fructose-1,6-biphosphate can be explained by the oscillations of fructose-6-phosphate caused by gluconeogenesis, rather than the allosteric property of the enzyme phosphofructokinase.

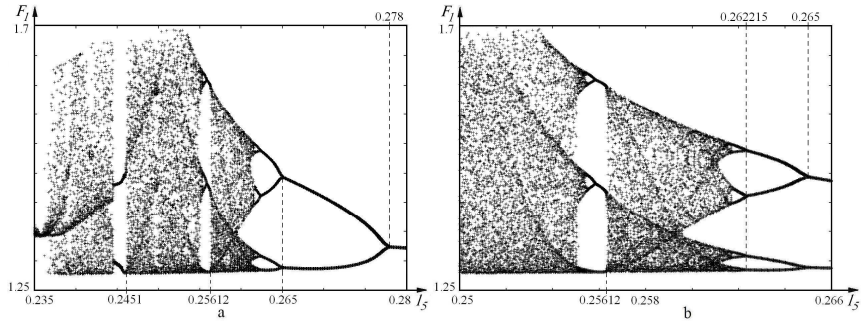

Fig. 2. Phase-parametric diagram of the system for the variable $F_1(t)$:
a - $l_5 \in (0.235, 0.28)$; b - $l_5 \in (0.25, 0.266)$.

The phase-parametric diagrams were constructed with the help of the cutting method. In the phase space, we took the cutting plane at $R_2 = 1.0$. This choice is explained by the symmetry of oscillations $F_1(t)$ relative to this point. At the cross of this plane by the trajectory, we fix the value of each variable. If a multiple periodic limiting cycle arises, we will observe a number of points on the plane, which coincide in the period. If a deterministic chaos arises, the points, where the trajectory crosses the plane, are located chaotically.

Considering the diagram from right to left, we may indicate that, at $l_5^j = 0.278$, the first bifurcation of the period doubling arises. Then at $l_5^{j+1} = 0.265$ and $l_5^{j+2} = 0.262215$, we see the second and third bifurcations, respectively. Further, the autooscillations transit in the chaotic mode due to the intermittence. The obtained sequence of bifurcations satisfies the relation

$$\lim_{t \to \infty} \frac{l_5^{j+1} - l_5^j}{l_5^{j+2} - l_5^{j+1}} \approx 4.668.$$

This number is very close to the universal Feigenbaum constant. The transition to the chaos has happened by the Feigenbaum scenario [35].

It is seen from Fig.2,a,b that, for $l_5 = 0.25612$ and $l_5 = 0.2451$, the periodicity windows appear. Instead of the chaotic modes, the periodic and quasiperiodic modes are established. The same periodicity windows are observed on smaller scales of the diagram. The similarity of diagrams on small and large scales means the fractal nature of the obtained cascade of bifurcations in the metabolic process created by gluconeogenesis.

As examples of the sequential doubling of a period of autoperiodic modes of the system by the Feigenbaum scenario, we present projections of the phase portraits of the corresponding regular attractors in Fig.3,a-c. In Fig.3,d-f, we show some regular attractors arising in the periodicity windows. For $l_5 = 0.256$, the 3-fold periodic mode $3 \cdot 2^0$ is formed. For $l_5 = 0.2556$, we observe the 5-fold mode. Then, as $l_5 = 0.245$, the 3-fold periodic cycle is formed again.

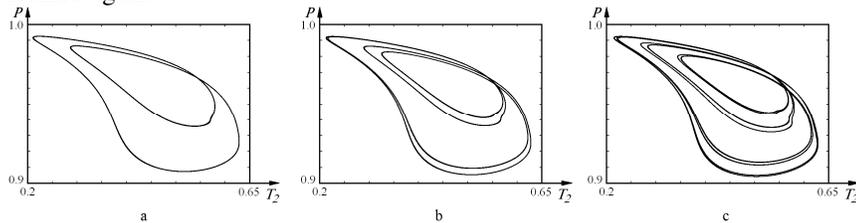

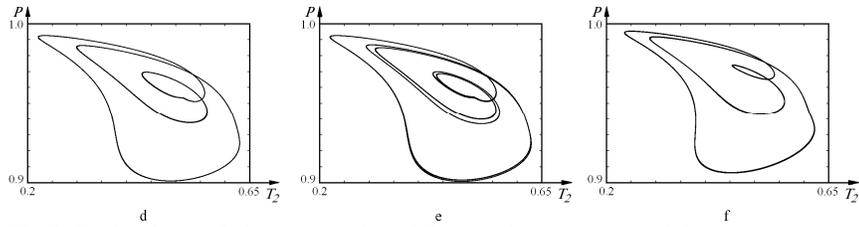

Fig.3. Projections of phase portraits of the regular attractors of the system:
a - $1 \cdot 2^1$, for $l_5 = 0{,}268$; b - $1 \cdot 2^2$, for $l_5 = 0.264$; c - $1 \cdot 2^4$, for $l_5 = 0.262$;
d - $3 \cdot 2^0$, for $l_5 = 0.256$; e - $5 \cdot 2^0$, for $l_5 = 0.2556$; and f - $3 \cdot 2^0$, for $l_5 = 0.245$.

In Fig. 4,a,b, we give projections of the strange attractor $2^x$ for $l_5 = 0.25$. The obtained chaotic mode is a strange attractor. It appears as a result of the formation of a funnel. In the funnel, there occurs the mixing of trajectories. At an arbitrarily small fluctuation, the periodic process becomes unstable, and the deterministic chaos arises.

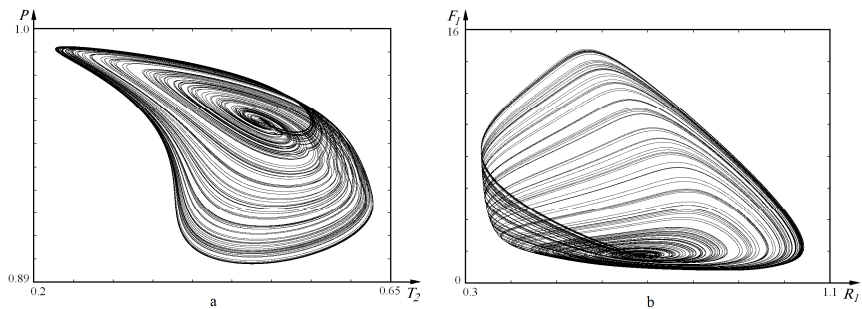

Fig.4. Projections of the phase portrait of the strange attractor $2^x$ for $l_5 = 0.25$: a – in the plane $(T_2, P)$, b – in the plane $(R_1, F_1)$.

In Fig.5,a,b, we present, as an example, the kinetics of autooscillations of some components of the metabolic process in a 1-fold mode for $l_5 = 0.3$ and in the chaotic mode for $l_5 = 0.25$. The synchronous autooscillations of fructose-6-phosphate and the inactive form $T_2$ of the allosteric enzyme phosphofructokinase are replaced by chaotic ones.

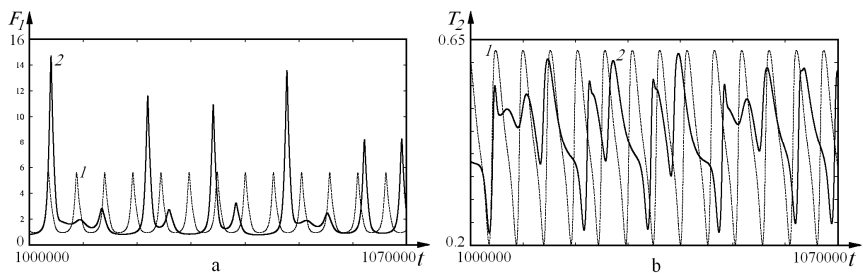

Fig.5. Kinetic curves of the variables: $F_1(t)$ - a and $T_2(t)$ - b in the 1-fold periodic mode for $l_5 = 0.3$ (1) and in the chaotic mode for $l_5 = 0.25$ (2).

While studying the phase-parametric diagrams in Fig.2,a,b, it is impossible beforehand to determine, for which values of parameter $l_5$ a multiple stable (quasistable) autoperiodic cycle or a strange attractor is formed.

For the unique identification of the type of the obtained attractors and for the determination of their stability, we calculated the complete spectra of Lyapunov indices $\lambda_1, \lambda_2, ... \lambda_{16}$ for chosen points and their sum $\Lambda = \sum_{j=1}^{16} \lambda_j$. The calculation was carried out by Benettin's algorithm with the orthogonalization of the perturbation vectors by the Gram--Schmidt method [13].

As a specific feature of the calculation of these indices, we mention the difficulty to calculate the perturbation vectors represented by $16 \times 16$ matrices on a personal computer.

Below in Table 1, we give several results of calculations of the complete spectrum of Lyapunov indices, as an example. For the purpose of clearness, we show only three first indices $\lambda_1 - \lambda_3$. The values of $\lambda_4 - \lambda_{16}$ and $\Lambda$ are omitted, since their values are not essential in this case. The numbers are rounded to the fifth decimal digit. For the strange attractors, we calculated the following indices, by using the data from Table 1. With the use of the Pesin theorem [36], we calculated the KS-entropy (Kolmogorov-Sinai entropy) and the Lyapunov index of a "horizon of predictability" [37]. The Lyapunov dimension of the fractality of strange attractors was found by the Kaplan--Yorke formula [38,39]:

By the calculated indices, we may judge about the difference in the geometric structures of the given strange attractors. For $l_5 = 0.25$, the KS-entropy takes the largest value $h = 0.00014$. In Fig.4,a, we present the projection of the given strange attractor. For comparison, we constructed the strange attractors for $l_5 = 0.26$ (Fig.6,a) and $l_5 = 0.237$ (Fig.6,b). Their KS-entropies are, respectively, 0.00008 and 0.00005. The comparison of the plots of the given strange attractors is supported by calculations. The trajectory of a strange attractor (Fig.4,a) is the most chaotic. It fills uniformly the whole projecton plane of the attractor. Two other attractors (Fig.6,a,b) have the own relevant regions of attraction of trajectories. The phase space is divided into the regions, which are visited by the trajectory more or less.

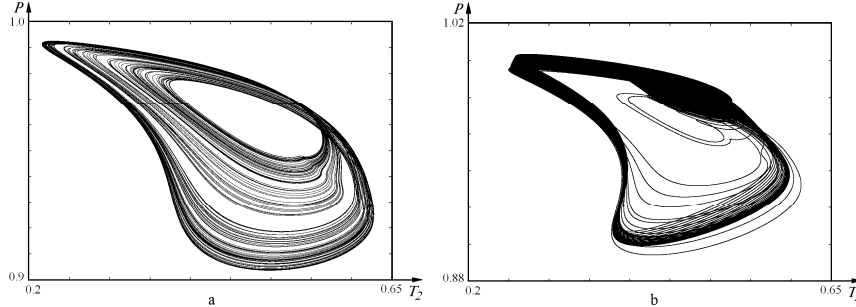

Fig.6. Projections of the phase portraits of the strange attractors $2^x$ in the plane $(T_2, P)$: a – for $l_5 = 0.26$ and b – for $l_5 = 0.237$.

Table 1. Lyapunov indices, KS-entropy, "horizon of predictability," and the Lyapunov dimension of the fractality of strange attractors calculated for various modes

| $l_5$ | Attractor | $\lambda_1$ | $\lambda_2$ | $\lambda_3$ | $h$ | $t_{min}$ | $D_{Fr}$ |
|---|---|---|---|---|---|---|---|
| 0.28 | $1 \cdot 2^0$ | .00000 | -.00008 | -.00010 | - | - | - |
| 0.264 | $1 \cdot 2^1$ | .00000 | -.00005 | -.00008 | - | - | - |
| 0.262 | $1 \cdot 2^2$ | .00000 | -.00005 | -.00009 | - | - | - |
| 0.26 | $2^x$ | .00008 | .00000 | -.00008 | .00008 | 12500 | 4 |
| 0.257 | $2^x$ | .00007 | .00000 | -.00008 | .00007 | 14285.7 | 3.9 |
| 0.256 | $3 \cdot 2^0$ | .00000 | -.00006 | -.00007 | - | - | - |
| 0.2556 | $5 \cdot 2^0$. | .00000 | -.00005 | -.00009 | - | - | - |

| 0.254 | $2^x$ | .00009 | .00000 | -.00009 | .00009 | 11111.1 | 4 |
| 0.252 | $2^x$ | .00012 | .00000 | -.00010 | .00012 | 8333.3 | 4.2 |
| 0.25 | $2^x$ | .00014 | .00000 | -.00009 | .00014 | 7142.9 | 4.6 |
| 0.248 | $2^x$ | .00009 | .00000 | -.00007 | .00009 | 11111.1 | 4.3 |
| 0.247 | $2^x$ | .00013 | .00000 | -.00010 | .00013 | 7692.2 | 4.3 |
| 0.2463 | $2^x$ | .00008 | .00000 | -.00011 | .00008 | 12500 | 3.7 |
| 0.245 | $3 \cdot 2^0$ | .00000 | -.00011 | -.00011 | - | - | - |
| 0.242 | $2^x$ | .00010 | .00000 | -.00008 | .00010 | 10000 | 4.25 |
| 0.24 | $2^x$ | .00006 | .00000 | -.00009 | .00006 | 16666.7 | 3.7 |
| 0.239 | $2^x$ | .00006 | .00000 | -.00010 | .00006 | 16666.7 | 3.6 |
| 0.238 | $2^x$. | .00011 | .00000 | -.00010 | .00011 | 9090.9 | 4.1 |
| 0.237 | $2^x$ | .00005 | .00000 | -.00008 | .00005 | 20000 | 3.6 |

The Lyapunov dimensions of the given strange attractors are changed analogously. We have, respectively: 4.6, 4, and 3.6. These values characterize generally the fractal dimension of the given attractors. If we separate small rectangular area on one of the phase curves in each of the given plots and increase them, we will see the geometric structures of the given strange attractors on small and large scales. Each arisen curve of the projection of a chaotic attractor is a source of formation of new curves. Moreover, the geometric regularity of construction of trajectories in the phase space is repeated for each strange attractor. In the given case, the best geometric self-similarity conserves in the presented strange attractors in the following sequence: Fig.4,a, Fig.6,a, and Fig.6,b.

The value of "horizon of predictability" $t_{min}$ for the modes presented in the table is the largest for $l_5 = 0.237$ (Fig.6,b). The narrow regions of attraction of the projection of the strange attractor correspond to the most predictable kinetics of the running metabolic process. From all metabolic chaotic modes, this mode is the mostly functionally stable for a cell.

The above-described study of the process of glycolysis-gluconeogenesis with the help of a change of the coefficient of positive feedback $l_5$ indicates that, in the given metabolic process under definite conditions, the autocatalysis arises. The value of $l_5$ determines the activity of gluconeogenesis on the section of the transformation of glucose-6-phosphate in glucose. This reaction is catalyzed by the enzyme glucose-6-phosphatase. This phosphatase is magnesium-dependent. If the magnesium balance is violated or some other factors come in play, the rate of this reaction varies. In addition, the absence of some coenzymes in a cell affects essentially also the rates of other enzymatic reactions, which can lead to the desynchronization of metabolic processes. As a result, the autooscillations arise in the metabolic process of glycolysis-gluconeogenesis. The autooscillations can be autoperiodic with various multiplicities or chaotic. Their appearance can influence the kinetics of the metabolic process in the whole cell and its physiological state.

## Conclusions

With the help of a mathematical model, we have studied the influence of gluconeogenesis on the metabolic process of glycolysis. The metabolic chain of glycolysis-gluconeogenesis is considered as a single dissipative system arisen as a result of the self-organization, i.e., as a product of the biochemical evolution in protobionts. The reasons for the appearance of autocatalysis in it are investigated. A phase-parametric diagram of autooscillatory modes depending on the activity of gluconeogenesis is constructed. We have determined the bifurcations of the doubling of a cycle according to the Feigenbaum scenario and have shown that, as a result of the intermittence, the aperiodic modes of strange attractors arise. The fractal nature of the calculated cascade of bifurcations is demonstrated. The strange attractors arising as a result of the formation of a mixing funnel are found. The complete spectra of Lyapunov indices for various modes are calculated. For strange attractors, we have calculated the KS-entropies, "horizons of predictability," and the Lyapunov dimensions of the fractality of attractors. The structure of a chaos of the given attractors and its influence on the stability of the metabolic process, adaptation, and functionality of a cell are studied. It is shown that a change of the cyclicity in the metabolic process in a cell can be caused by the violation of the magnesium balance in it or the absence of some coenzymes. The obtained results allow one to study the influence of gluconeogenesis on the self-organization of the metabolic process in a cell and to find the reasons for a change of its physiological state.

*The work is supported by project* N 0112U000056 *of the National Academy of Sciences of Ukraine.*